\documentclass[pra,twocolumn,showpacs,superscriptaddress,floatfix]{revtex4}
\usepackage{color}
\usepackage{hyperref}  
\usepackage{graphicx}
\usepackage{amsfonts}

\let\a=\alpha \let\b=\beta    \let\d=\delta \let\e=\varepsilon
     \let\th=\theta  \let\l=\lambda
\let\m=\mu    \let\n=\nu         \let\p=\pi    \let\r=\rho
\let\s=\sigma \let\t=\tau    
     
\let\G=\Gamma

\font\tenmib=cmmib10\font\sevenmib=cmmib7\font\fivemib=cmmib5%
\mathchardef\Bl   = "0515  

\def\Bl   {{\mbox{\boldmath$ \lambda$}}}

\def\Bx   {{\mbox{\boldmath$ \xi$}}}

\def\BDpr {{\mbox{\boldmath$ \partial$}}}

\def\eqalign#1{\null\,\vcenter{\openup\jot
  \ialign{\strut\hfil$\displaystyle{##}$&$\displaystyle{{}##}$\hfil
      \crcr#1\crcr}}\,}

\def\CC{{\mathcal C}}\def\FF{{\mathcal F}}
\def\EE{{\mathcal E}}\def\DD{{\mathcal D}}\def\TT{{\mathcal T}}

\def\uu{{\V u}}\def\kk{{\V k}}\def\xx{{\Bx}}
\def\T#1{{#1_{\kern-3pt\lower7pt\hbox{$\widetilde{}$}}\kern3pt}}
\def\Tdpr{{\T{\BDpr}}}
\def\uU{{\T {\bf u}}}\def\Ff{{\T{\bf F}}}
\def\Gg{{\V g}}
\def\ie{{\it i.e.\ }}
\def\dpr{{\partial}}
\def\defi{{\buildrel def\over=}}
\newdimen\xshift \newdimen\xwidth \newdimen\yshift \newdimen\ywidth

\def\ins#1#2#3{\vbox to0pt{\kern-#2pt\hbox{\kern#1pt #3}\vss}\nointerlineskip}

\def\eqfig#1#2#3#4#5{
\par\xwidth=#1pt \xshift=\hsize \advance\xshift
by-\xwidth \divide\xshift by 2
\yshift=#2pt \divide\yshift by 2
{\hglue\xshift \vbox to #2pt{\vfil
#3 \includegraphics{#4.eps}
}\hfill\raise\yshift\hbox{#5}}}

\def\V#1{{\bf #1}}
\def\lis#1{{\overline#1}}

\def\tende#1{\,\vtop{\ialign{##\crcr\rightarrowfill\crcr
 \noalign{\kern-1pt\nointerlineskip} \hskip3.pt${\scriptstyle
   #1}$\hskip3.pt\crcr}}\,}
\def\eg{{\it e.g.\ }}
\def\cfr{{\it c.f.r.} }
\def\0{\noindent}
\def\Eq#1{\label{#1}}
\def\equ#1{(\ref{#1})}

\font\titolo=cmbx12%
\def\iniz{\setcounter{equation}{0}}
\def\be{\begin{equation}}\def\ee{\end{equation}}

\usepackage{fancyhdr}
\def\alert#1{{\color{ired}#1}}
\definecolor{iblue}{RGB}{65,105,225}
\definecolor{ired}{RGB}{220,20,60}
\definecolor{igreen}{RGB}{50,205,50}
\definecolor{ipurple}{RGB}{75,0,130}
\definecolor{iochre}{RGB}{218,165,32}
\definecolor{iteal}{RGB}{51,204,204}
\definecolor{imauve}{RGB}{204,51,153}

\iniz
\begin{document}
\let\titolo=\bf
\alert{\centerline{\titolo Reversible viscosity and
    Navier--Stokes fluids}}\vskip1mm

\centerline{\bf Giovanni Gallavotti} \centerline{\today}

{\vskip3mm}
\noindent {\bf Abstract}: {\it Exploring the possibility of describing a
  fluid flow via a time-reversible equation and its relevance for the
  fluctuations statistics in stationary turbulent (or laminar)
  incompressible Navier-Stokes flows.}  {\vskip3mm}

\def\SEC{Introduction}
\section{\SEC}
\label{sec1}

Studies on non equilibrium statistical mechanics progressed after the
introduction of thermostats, \cite{EM990}. Finite thermostats have not only
permitted a new series of simulations of many particle systems, but have
been essential to clarify that {\it irreversibility} and dissipation
{\it should not} be identified.

Adopting the terminology of \cite{FV963} it is convenient to distinguish
the finite system of interest, \ie particles forming the {\it test system}
in a container $\CC_0$, from the thermostats.  The thermostats
$\TT_1,\TT_2,\ldots$ are also particle systems, forming the {\it
  interaction systems}, acting on the test systems: they are in infinite
containers and, {\it asymptotically at infinity}, are always supposed in
equilibrium states with given densities $\r_1,\r_2,\ldots$
and temperatures $T_1,T_2,\ldots$. 

The thermostats particles in each thermostat may interact with each other
and with the particles of the test system {\it but not directly} with the
particles of the other thermostats. The test system and the interaction
systems, together, form a Hamiltonian system (classical or quantum) that can
be symbolically illustrated as in Fig.0:

\kern-.5cm
\eqfig{100}{100}{}{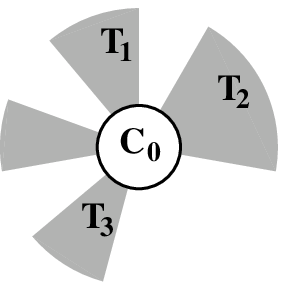}{}

\0{\small Fig.0: The ``{\it test}'' system are particles enclosed in $\CC_0$
  while the external $\CC_j$ systems are thermostats or,
  following the terminology of Feynman--Vernon, \cite{FV963},
  {\it ``interaction''} systems.}

\*

Finite thermostats have been introduced recently and fulfill the main
function of replacing, \cite{EM990}, the above test systems and ``perfect
thermostats'', consisting of infinite systems of particles in a state in a
well defined equilibrium state at infinity, with finite systems suitable
for simulations.

The perfect thermostats, being infinite, are not
suited in simulations, while the finite ones have the drawback that their
equations of motion contain ``unphysical forces''.

The basic idea is that, asymptotically \eg for large number of
particles (``thermodynamic limit''), most statistical properties of the
``test'' system do not depend on the particular thermostat model but only
on its equilibrum parameters defined at infinity.

Several finite thermostats employed in simulations are governed by
reversible equations of motion: denoting $u\to S_t u, t\in R$ the time
evolution of a point $u$ in phase space $\FF$, this means that the map
$u\to Iu$ in which all velocities in $u$ are reversed is such that $S_tI=I
S_{-t}$, so that if $u(t), t\in R$ is a possible solution of the equations of
motion also $I\,u(-t), t\in R$ is a possible one.

If $u$ describes the state of a system in which dissipation occurs, \ie in
which external forces perform work on the test subsystem, it might be
thought that, unless the interaction systems are infinite, the motion is
not reversible: this has been clearly shown to be not true by the many
simulations performed since the early '80s, reviewed in \cite{EM990}.
And the simulations have added evidence that the same physical
phenomenon occurring in the test system is largely independent of several
(appropriate) realizations of thermostat models (reversible or not).

A remarkable instance is an example of a system of particles interacting with a
single thermostat at temperature $\b^{-1}=T$ which has a stationary state
described by a probability distribution $\m(du)$ which is different from
the canonical distribution (say) but which is nevertheless equivalent to it
in the sense, \cite{Ga999}, of the theory of ensembles, \ie in the
thermodynamic limit, see \cite{EM990}.

In the different context of turbulence theory a similar example can be
found in the simulation in \cite{SJ993}: where viscosity is set $=0$ but
``unphysical forces'' are introduced to constrain the energy value on each
``energy shell'' to fulfill the OK ``$\frac53$ law''. The stationary
distribution of the velocity field for many observables, \eg the large scale
velocity components, remains the same as in the viscous unconstrained
system and in the reversible new one, at very large Reynolds number.

Then one is led to think that the root of the equivalence between very
different equations of motion for the same physical system lies in the
fundamental microscopic reversibility of the equations of motion,
\cite{Ga996b,Ga997b}, and to a precise formulation of the ``conjecture''
that ``{\it in microscopically reversible (chaotic) systems time reversal
  symmetry cannot be spontaneously broken, but only phenomenologically
  so}'' and a program to test it, was proposed, \cite{Ga998}. The program
has been followed so far in a few works, \cite{GRS004,GL014}, with results
apparently not always satisfactory \cite{RM007}.

Here, after a general discussion of the conjecture and its precise
formulation, several tests will be proposed, on the statistical properties of
the stationary states of the $2D$ incompressible Navier-Stokes equation,
and performed with results described in some detail.

\def\SEC{Irreversible and Reversible ODE'{\small s}}
\section{\SEC}
\label{sec2}

More generally an ODE $\dot x=h(x)$ on the ``phase space'' $R^N$ has a {\it
time reversal symmetry} $I$ if the solution operator $x\to S_tx, x\in
R^N$, and the map $I$ are such that $I^2=1, \,S_tI=IS_{-t}$. 

Non trivial
examples are provided, as mentioned, by many Hamiltonian equations, but
there are also interesting examples not immediately related to Hamiltonian
systems, as the equations of the form $\dot x_j= f_j(x),\,
\n>0$,\\$j=1,\ldots,N$ with $f_j(x)=f_j(-x)$, like the Lorenz96 model at
$\n=0$:
\be\dot x_j=x_{j-1}(x_{j+1}-x_{j-2})+F-\n x_j, \Eq{e2.1}\ee
with $F=const$ and periodic b.c. $x_0=x_{N}$.

Another example is provided by the {\rm GOY\it shell
model}, \cite{BPPV993,Bi003}, given by: 
\be\eqalign{ \dot u_n=&-\n k_n^2 u_n + g\d_{n,4}\cr
  +ik_n&\big( -\frac14\lis u_{n-1}\lis u_{n+1}+\lis u_{n+1}\lis
  u_{n+2}-\frac18\lis u_{n-1}\lis u_{n-2}\big)\cr}\Eq{e2.2}\ee
where $k_n=2^n$, $u_n=u_{n,1}+iu_{n,2}$, with $u_n=0$ for $n=-1,0$ or
$n>N$, if $\n=0$.

A reversible equation often evolves initial data $x$ into functions $x(t)$
which are unbounded as $t\to\infty$. The case of Hamiltonian systems with
bounded energy surfaces are an important exception. Therefore, particularly
in problems dealing with stationary states in chaotic systems, the
equations contain additional terms which arise by taking into account that
the systems under study are also subject to stabilizing mechanisms
forcing motions to be confined to some sphere in phase space.

A typical additional term is $-\n x_j$ or $-\n (Lx)_j$ with $\n>0$ and $L$
a positive defined matrix: such extra terms are often introduced
empirically. This is the case in the above two examples. And they can be
thought as empirical realizations of the action of thermostats acting on
the systems.

At this point it is necessary to distinguish the models in which
\*

\0(1) the equations $\dot x=h(x)-\n Lx$ arise, possibly in some limit case,
from a system of particles, as the one of the Feynman-Vernon system in
Sec.1, Fig.0, or \\
\0(2) the equations are not directly related to a fundamental microscopic
description of the system.
\*

The above Lorenz96 and GOY models are examples of the second case, while the
Navier-Stokes equations, since the beginning, were considered macroscopic
manifestations of particles interacting via Newtonian forces,
\cite[Eq.(128)]{Ma867-b}.

The success of the simulations using artificial thermostat forces with finite
thermostats and the independence of the results from the particular choice
of the thermostats used to contain energy growth in nonequilibrium, \cite{EM990}, induces
to think that there might be alternative ways to describe the same systems
via equations that maintain the time reversal symmetry shown by the non
thermostatted equations. A first proposal that seems natural is the
following.

Consider an equation
\be \dot x=h(x)-\n Lx,\  {\rm with}\ h(x)=h(-x)\Eq{e2.3}\ee
time reversible if $\n=0$, for the time reversal $Ix=-x$; suppose that $|x\cdot
h(x)|\le \G (x\cdot Lx)$. Then the motions will be asymptotically confined, if $\n>0$,
to the ellipsoid $(x\cdot Lx)\le \frac{G}\n$ and the system will be able to 
reach a stationary state, \ie an invariant probability distribution
$\m^C_{\frac1\n}$ of the phase space points. Frequently, if $\n$ is small
enough, the motions will be chaotic and there will be a unique stationary
distribution, the ``SRB distribution'', \cite{Ru995}.

The family of stationary distributions forms what will be called the ``{\it
  viscosity ensemble}'' $\FF^C$ whose elements are parameterized by $\n$
(and possibly by an index distingushing the extremal distributions which
can be reached as stationary states, for the same $\n$, from different
initial data); then consider the new equation

\be \dot x=h(x)-\a(x) Lx,\qquad \a(x)=\frac{(Lx\cdot h(x))}{(Lx\cdot Lx)}
\Eq{e2.4}\ee
where $\a$ has been determined so that the observable $\DD(x)\defi(x\cdot
Lx)$ is an exact constant of motion. For each choice of the parameter $\EE$
the evolution will determine a family $\m^M_\EE$ of stationary probability
distributions parameterized by the value $\EE$ that  $\DD$ takes on the
initial $x$ generating the distribution. The collection $\FF^M$ of such
distributions will be called ``{\it reversible viscosity ensemble}''
because the distributions are stationary states for Eq.\equ{e2.4} which is
reversible (for $Ix=-x$).

Also in this case if $\EE$ is large the evolution Eq.\equ{e2.4} is likely
to be chaotic and for each such $\EE$ the distribution $\m^M_\EE$ is
unique: {\it if not} extra parameter needs to be introduced the identify
each of the extremal ones.

Suppose for simplicity that $\frac1\n,\EE$ are large enough and the
stationary states $\m^C_{\frac1\n},\m^M_\EE$ are unique. Then say that 
$\m^C_{\frac1\n}$ and $\m^M_\EE$ are {\it correspondent} if
\be \m^M_\EE(\a)=\frac1\n, \qquad {\rm or\ if}\qquad
\m^C_{\frac1\n}(\DD)=\EE\Eq{e2.5}\ee
Then the following proposal appears in \cite{Ga996b,Ga997b} about the
properties of the fluctuations of ``K-local observables'', \ie of observables
$F(x)$ depending only on the coordinates $x_i$ with $i<K$ 
\vglue3mm

\0{\it If $\frac1\n$ and $\EE$ are large enough so that the motions
  generated by the equations Eq.\equ{e2.3},\equ{e2.4} are chaotic, \eg
  satisfy the ``Chaotic hypothesis'', \cite{GC995, GC995b}, then
  corresponding distributions $\m^C_{\frac1\n},\m^M_\EE$ give the same
  distribution to the fluctuations of a given K-local observable $F$
  in the sense that
  \be\m^M_\EE( F)=\m^C_{\frac1\n}(F)(1+o(F,\n))\Eq{e2.6}\ee
  with $o(F,\n)\tende{\frac1\n\to\infty}0$.}
\*

There have been a few attempts to check this idea, \cite{GRS004,GL014} and
more recently in \cite{De017}. 

\def\SEC{Reversible viscosity}
\section{\SEC}
\label{sec3}
\iniz

The ideas of the preceding section will next be studied in the case of the
Navier-Stokes equation. This is particularly interesting becasuse the
equation can be formally derived as an equation describing the macroscopic
evolution
of microscopic Newtonian particles (\ie point massses interacting
via a short range force), \cite{Ma867-b}. Hence the equation belongs to the
rather special case (1) in Sec.2.

The incompressible Navier-stokes equations with viscosity $\n$ for a
velocity field $\V v(\xx,t)$ in a periodic container of size $L$ and with a
forcing $\V F=F\Gg$ acting on large scale, \ie with Fourier components $\V
F_\kk\ne0$ only for a few $|\kk|$. {\it To fix the ideas} in $2$ dimensions
choose $F_\kk\ne0$ only for the single mode $\kk=\pm\frac{2\p}L
(2,-1)$ with $||\V F||_2=F$ (\ie $\Gg_{\pm\kk}=\frac{e^{\pm i\th}}{\sqrt2}$
for some phase
$\th$).

The equations can be written in dimensionless form:
introduce rescaling parameters
$V,T$ for velocity and time, and write $\T{\V v}(\xx,\t)=V
\uU(\xx/L,\t/T)$. 
Define $V=(FL)^{\frac12},\ T=(\frac{L}F)^{\frac12}$ and fix
$\frac{TV}{L}=1$ and $\frac{FT}V=1$; then the equation for $\uu(\V x,t)$
can be written as, ``I-NS'':
\be
\dot \uU+(\uu\cdot\BDpr) \uu=\frac1R\Delta \uU + \Gg - \BDpr p ~, \quad
\BDpr \cdot \uu = 0
\Eq{e3.1}\ee
where $R\equiv\frac{LV}\n \equiv (\frac{F L^3}{\n^2})^{\frac12}$ and $p$ is
the pressure. In this way the inverse of the viscosity can be identified
with the dimensionless parameter $R$, ``Reynolds number''.

The units for $L,F$ will be fixed so that $F=1$ and $L=2\p$: hence the
modes $\kk$ will be pairs of integers $\kk=(k_1,k_2)$.
The reality conditions
  $\uu_\kk=\lis\uu_{-\kk}, F_\kk=\lis F_{-\kk}$ implies that only the
components with
\be\kk=(k_1,k_2)\in I^+ \defi \{k_1> 0\ {\rm
  or}\ k_1=0,k_2\ge0\}\Eq{e3.2}\ee
are independent components (and it is assumed that $\uu_{\V0}=0$).

We shall consider the case of $2$ dimensional incompressible fluids to
avoid the problem that the $3$ dimensional equations have not yet been
proved to admit a (classical or even just constructive) solution. In spite
of this, below, the $3$ dimensional case will also be commented and
essentially everything that will be presented in the $2$ dimensional case
{\it turns out also relevant in $3$ dimensions}.

Proceding as in sec.2, define the family $\FF^C$ of stationary
probability distribution $\m^{\bf C}_R(d\uu)$ on the fields $\uu$
corresponding to the {\it stationary state} for the Eq.\equ{e3.1}.

Consider, {\it alternatively}, the equation
(reversible for the symmetry
$I\uu=-\uu$), ``R-NS'':
\be
\dot \uU+(\uu\cdot\BDpr) \uU=\a(\uU)\Delta \uU + \Ff - \Tdpr p ~, \quad
\Tdpr \cdot \uU = 0
\Eq{e3.3}\ee
in which the viscosity $\n=\frac1R$, \cfr Eq.\equ{e3.1}, is replaced by the
multiplier $\a(\uu)$ which is fixed so that
\be \DD(\uu)=\int |\Tdpr \uu(\V x)|^2 d\V x={\rm exact\ const.\ of\
  motion}\Eq{e3.4}
\ee
Therefore, if the space dimension is $2$, the multiplier $\a(\uu)$ will be
expressed, in terms of the Fourier transform $\uu_\kk$ (defined via $\uu(\V
x)=\sum_{\kk} e^{2\p i \kk\cdot{\V x}} \uu_\kk$) as:
\be\kern-3mm\a(\uu)= \frac{\sum_\kk \kk^2\lis {\Gg}_\kk\cdot \uu_\kk}
    {\sum_\kk \kk^4 |\uu_\kk|^2}\equiv
\frac{\sum_{\kk\in I^+} \kk^2 (g^r_\kk \uu^r_\kk+g^i_\kk \uu^i_\kk)}
     {2\sum_{\kk\in I^+} \kk^4 |\uu_\kk|^2}
\Eq{e3.5}\ee
and the stationary distribution for Eq.\equ{e3.5} with the value
of $\DD(\uu)$ fixed to $\EE$, will be denoted
$\m^{\bf M}_{\EE}(d\uu)$.

The collection of all stationary distributions $\m^{\bf C}_R$ as $R$ varies
and of all stationary distributions $\m^{\bf M}_\EE$ as $\EE$ varies will
be denoted $\FF^C$ and $\FF^M$ and called {\it viscosity ensemble}, as in
sec.2, and, respectively, {\it enstrophy ensemble}.

Call {\it $K$-local} an observable $f(\V u)$ which depends on the finite
number of components $\uu_\kk$ with $|\kk|<K$, of the velocity field; then
in the above cases the {\it conjecture} proposed in \cite{Ga996b,Ga997b}
becomes \*

\0{\it In the limit of large Reynolds number the distribution $\m^{C}_R$
  attributes to any given $K$-local observable $f(\V u)$ the same average, in
  the sense of Eq.\equ{e2.6} with $R\equiv\frac1\n$, as
  the distribution $\m^{\bf M}_{\EE}$ if
  \be \EE=\int \m^{\bf
    C}_R(d\uu) \DD(\uu)\Eq{e3.6}\ee
}
\*

\0{\it Remarks:}
(1) The size of $R$ might (of course ?, see however Sec.4)
depend on the observable
$f$, \ie on how many Fourier modes are needed to define $f$.
\\
(2) Therefore locality in Fourier space is here analogous to locality in
space in the equivalence between equilibrium ensembles.
\\
(3) The notations $\m^{\bf C}_R,\m^{\bf M}_{\EE}$ have been used to evoke
  the analogy of the equivalence between canonical and microcanonical
  ensembles in equilibrium statistical mechanics: the viscosity ensemble
  can be likened to the canonical ensemble, with the viscosity $\n=\frac1R$
  corresponding to $\b$, and the enstrophy ensemble to the microcanonical
  one, with the enstrophy corresponding to the total energy.
  \\
(4) The equivalence has roots in the {\it chaotic hypothesis},
  \cite{GC995b}: if the motion is sufficiently chaotic, as expected if $R$
  or $\EE$ are large, \cite{RT971,Ru995}, the multiplier $\a(\uu)$
  fluctuates in time and the conjecture is based on a possible ``{\it
    self-averaging}'' of $\a$ implying homogeneization of $\a(\uu)$ in
  Eq.\equ{e3.3} to a constant value, namely $\n=\frac1R$.
  \\
(5) the latter remark, if $\m^C_R$ is equivalent to $\m^M_\EE$ (\eg if
  $\m^C_R(\DD)=\EE$, see Eq.\equ{e3.6},\equ{e2.5}), leads to expect a relation like:
  \be \m^{\bf M}_R(\a)=\frac1R(1+o(\frac1R)),\Eq{e3.7}\ee\\
(6) The property $\a(\uu)=-\a(-\uu)$ implies that the evolution defined by
  Eq.\equ{e3.3} is {\it time reversible}, so that $\a(\uu)$ can be called
  ``{\it reversible viscosity}''.

\def\SEC{Regularization}
\section{\SEC}
\label{sec4}
\iniz

In Eq.\equ{e2.6},\equ{e3.7} the question on how large should $R$
be for equivalence is implicitly raised. An answer, which may become
relevant in  simulations, that it would be
interesting to investigate, is that the equivalence might hold much more
generally, at least in the cases (1) in Sec.2 above: therefore for the
Navier Stokes equations in dimension $2$ (and $3$, see below).

The Navier-Stokes equation in $2D$ is known to admit unique evolution of
smooth initial data, \cite{Ga002}. The same question has not yet been
studied for the reversible viscosity case. In {\it both cases}, however,
simulations impose that the field $\uu$ must be represented by a finite
number of data, \ie it must be ``regularized'', to use the language of
field theory, \cite{Ga985}.

Here the regularization will simply be enforced by considering
Eq.\equ{e3.1},\equ{e3.3} with fields with $\uu_\kk\ne 0$ only if $\kk\in
I_N\defi\{|k_j|\le N\}$. Consequently all statements will depend on the
cut-off value $N$. In particular the conjecture of equivalence will have to
be studied also as a function of $N$ and for a fixed local observable.

Pursuing the analogy wih equilibrium statistical mechanics, SM, of a system
with energy $E$, temperature $\b^{-1}$ and observables localized in a
volume $V_0$, mentioned above, consider

\*

\0{\it(a) the cut-off $N$ as analogous to the total volume in SM, 
\\
(b) $K$--local observables (defined before Eq.\equ{e2.5}) as analogous to
the observables localized in a volume $V_0=K$ in SM
\\
(c) the enstrophy $\DD(\uu)$ as analogous to the energy in SM
}
\*

Furthermore the incompressible Navier Stokes equations (as well as the
Euler equations or the more general transport equations) can be regarded, if
$N=\infty$, as macroscopic versions of the atomic motion: the latter is
certainly reversible (if appropriately described together with the external
interactions) and essentially always strongly chaotic.

Therefore, for $N=\infty$ and at least for $2$ dimensions, no matter
whether $R$ is small or large, the equivalence should not only remain valid
but could hold in stronger form. Let $\m^M_{\EE,N},\m^C_{R,N}$ be the
stationary distributions for the regularized Navier-Stokes equations, then

\*

\0{\it Fixed $K$ let $F$ be a $K$-local observable;
  suppose that the equivalence condition $\m^C_{R,N}(\DD)=\EE$ (or
  $\m^M_{\EE,N}(\a)=\frac1R$)  holds,
  then:
  \be\eqalign{(a)&
    \m^C_{R}=  \lim_{N\to\infty}\m^C_{R,N},\ 
    \m^M_{\EE}=\lim_{N\to\infty}\m^M_{\EE,N}\ {\rm exist}\cr
  (b)&\ \m^C_{R}(F)=\m^M_{\EE}(F),\kern1.5cm {\rm for\ all}\quad R,\EE\cr}
  \Eq{e4.1}\ee
}

\0{\it Remarks:} (1) The statement is much closer in spirit to the familiar
thermodynamic limit equivalence between canonical and microcanonical
ensembles.
\\
(2) Since the basis is that the microscopic motions that generate the
Navier-Stokes equations are chaotic and reversible the limit $N\to\infty$
is essential.
\\
(3) The full Navier stokes equations at {\it low Reynolds number} admit,
for the same $R$, fixed point solutions, periodic solutions or even
coexisting chaotic solutions, \cite{FT985, FTZ984}, the condition of
equivalence must be interpreted as meaning that when there are several
coexisting stationary {\it ergodic} distributions then there is a
one-to-one correspondence between the ones that in the two ensembles
$\FF^C,\FF^M$ obey the equivalence condition and the averages of local
observables obey Eq.\equ{e4.1}.
\\
(4) The possibility of coexisting stationary distributions is analogous to the
phase coexistence in equilibrium statistical mechanics (and in that case too
the equivalence can hold only in the thermodynamic limit).
\\
(5) It is remarkable that above conjecture {\it really deals only with the
regularized equations}: therefore it makes sense irrespective of whether the
non regularized equations dimensionality is $2$ or $3$.\\ Of course in the
$3$--dimensional equation the $\a(\uu)$ has a somewhat different form,
\cite{Ga002}; furthermore in the developed turbulence regimes, {\it in
  dimension $3$}, the picture may become simpler: this is so because of the
     {\it natural cut-off due to the OK41 $\frac53$-law}: namely $|k_j|\le
     N=R^{\frac34_\e}$, $\e>0$, \cite{Ga002}.  \\
(6) The equivalence also suggests that there might be even some relation
     between the ``{\it $T$-local Lyapunov exponents}'' of pairs of
     equivalent distributions. Here $T$--local exponents are defined via
     the Jacobian matrix $M_T(\uu)=\dpr S_T(\uu)$ and its
     $RU$-decomposition: they are the averages of the diagonal elements
     $\l_j(\uu)$ of the $R$-matrix over $T$ time steps of integration,
     \cite{BGGS980a}.  Although the ``local exponents'' cannot be
     considered to be among the $K$-local observables it is certainly worth
     to compare the two spectra.
\\ (7) A suggestion emerges that it would be interesting to study the
R-NS equations with $\a(\uu)$ {\it replaced by a stochastic process} like a
white noise centered at $\frac1R$ with the reversibility taken into account
by imposing the width of the fluctuations to be also $\frac1R$, as required
by the fluctuation relation, see below. As $R$ varies stationary states
describe a new ensemble which could be equivalent to $\EE^C$ in the sense
of the conjecture.  \\
(8) A heuristic comment: if the {\it Chaotic hypothesis}, \cite{GC995b}, is
assumed for the evolution in the regularized equations the {\it fluctuation
  relation}, see below, should also hold, thus yielding a prediction on the
large fluctuations of the observable ``divergence of the equations of
motion'' $\s_N(\uu)$ in the distributions $\m^{\bf M}_{\EE,N}$ which, in
the $2$-dimensional case, is:
\def\hh{{\V h}}
\be
\kern-3mm
\s_N(\uu)=-\frac{\sum_{\hh\in I_N} \hh^4 {\rm Re}(\lis \Gg_\hh\cdot\uu_\hh)
  -
2\a E_6}{E_4}-\a\sum_{\hh\in I_N} \hh^2\Eq{e4.2}\ee
where $I_N\defi\{|k_j|\}\le N, \,E_{2m}=\sum_{\hh\in I_N} \hh^{2m}\,
|\uu_\hh|^2$ which follows, if $g_\hh\defi g^r_\hh+i\,g^i_\hh$,  from
\be
\frac{\dpr\a}{\dpr \uu^b_\hh}=\frac{\hh^2 g^b_\hh}{E_4}
-2\a \frac{ \hh^4 \uu^b_\hh}{E_4},\quad b=r,i
\Eq{e4.3}\ee
Notice that the cut-off $N$ is essential to define $\s_N(\uu)$ as the last
(and main) term in Eq.\equ{e4.2} would be, otherwise, infinite.

If $\s_{N,+}$ is the infinite time average of $\s_N(S_t\uu)$, \ie
  $\s_{+,N}\equiv \int \m^{M}_R(d\uu)\,\s_N(\uu)$ and if
  $p_\t(\uu)=\frac1\t\int_0^\t \frac{\s_N(S_t\uu)}{\s_{N,+}} dt$ then the
  variable
  $p_\t(\uu)$ satisfies the fluctuation relation in $\m^{M}_{En,N}$ if,
  asymptotically as $\t\to\infty$,
  \be \frac{\m^{\bf M}_{En,N}(p_\t(\uu)\sim p)}
      {\m^{\bf M}_{En,N}(p_\t(\uu)\sim -p)}
=e^{p  \s_{N,+}\t+ o(\t)}\Eq{e4.4}\ee
The average $\s_{N,+}$ becomes infinite in the limit $N\to\infty$: which
implies that the probability of $|p-1|>\e$ tends to $0$ (exponentially in
$N^4$, \ie proportionally to $ \e^2\sum_{|\kk|<N} \kk^2$) so that the
reversible viscosity (proportional to $\a\sim \frac{\s}{\s_+}$) will have
probability tending to $0$ as $N\to\infty$ (if
the large deviation function has a quadratic maximum at
$p=1$ or faster if the maximum is steeper). Large fluctuations of the
reversible viscosity away from $\frac1R$ are still possible if $N< \infty$
but not observable, \cite[Eq.(5.6.3)]{Ga013b}.

Some of the questions raised in the remarks in the above sections will now
be analyzed in a series of simulations in the next Appendix. They are very
preliminary tests and are meant just to propose
tests to realize in the future to test validity, dependence/stability of
the results as $N,R$ vary. Source-codes (in progress) available on request.

\vglue-3cm

\def\SEC{Appendix: Reversible viscosity and Reynolds number}
\section{\SEC}
\label{sec5}
\iniz

We first analyze the evolution and distribution of the
reversible viscosity $\a(\uu)$ defined in Eq.\equ{e3.5} considered as an
observable for the evolution Eq.\equ{e3.1}, \ie for the {\it irreversible
  NS2D evolution}.

Consider the NS2D with regularization $(2N+1)\times (2N+1)$. For $N=3$
a simulation gives the running

\eqfig{240}{180}{}{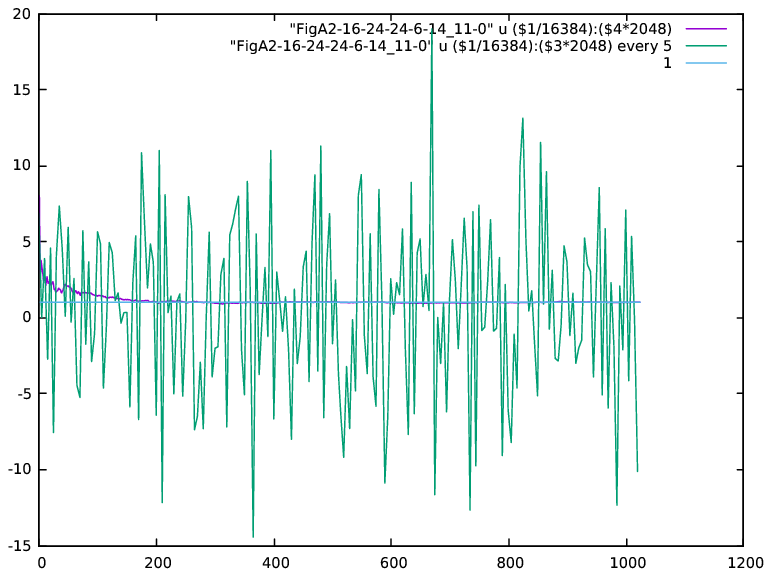}{}{%
  \hbox{\tiny FigA2-16-24-24-6-14\_11-0}}

\0{\small Fig.1: the modes are in the $7\times7$ box centered at the
  origin, corresponding to a cut-off $N=3$; the Reynolds number is
  $R=2^{11}$; the time step is $2^{-14}$ and
  the time axis is in units of $2^{14}$ (\ie the evolution history is
  obtained via $2^{24}$ time steps). }

\*

\0average of the value of $R\a(\uu)$ (drawn every $5$ data to avoid a too
dense a figure), the actual fluctuating values of $R\a(\uu)$ and the straight
line at quota $1$.  It shows that $R\a(\uu)$ fluctuates strongly, yet
$R\a(\uu)$ averages to a value close ($\sim2\%$) to $1$, \ie $\a(\uu)$
averages to the viscosity value: the analogy, mentioned earlier in
Eq.\equ{e3.7}, with equilibrium thermodynamics would suggest checking that
at large $R$, $\m^C_{R,N}(\a)=\frac1R (1+o_{R,N})$ with $o_{R,N}$ small. A
check is also necessary because $\a(\uu)$ is not a $K$-local observable.

The same data considered in Fig.1 for $R=2014$ and $2^{26}$ integration steps
of size $2^{-15}$ drawn every $10\cdot 2^{15}$ yield:

\eqfig{240}{175}{}{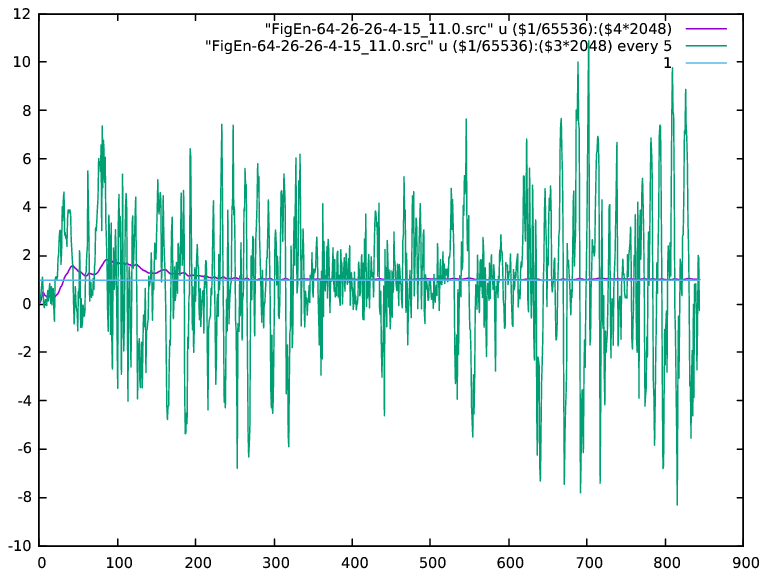}{}
\raise 3mm\hbox{\kern5.5cm\tiny FigA2-64-26-26-4-15-11.0}

\vglue-.2cm \0{\small Fig.2: At $960$ modes and $R=2048$: the evolution of
  the observable ``{\it reversible viscosity}'', \ie $\a(\uu)$ in
  Eq.\equ{e3.5} in the I-NS: the time average of $\a$ should be
  $\frac1R(1+o(\frac1R))$. Represents the fluctuating values of $\a$ every
  $5\cdot 2^{16}$ integration steps; the middle line is the
  running average of $\a$ and it {\it is
    close} to $\frac1R$ (horiz. line).}

\*
It would be interesting to present a few more recent results on the
closeness of the Lyapunov spectra of the R-NS and
I-NS but the analysis requires further work.

\0{\bf Acknowledgements: \it This is an extended version of part of my talk
  at and includes only the material prepared to propose simulations to the
  attending postdocs during my stay at the Institut Henri Poincar\'e -
  Centre Emile Borel during the trimester  {\sl Stochastic Dynamics Out of
  Equilibrium}. I am grateful to the organizers for the support and
  hospitality and also for the possibility of starting and performing the
  presented simulations on the IHP computer cluster; I thank also
  L. Biferale for providing computer facilities to improve the graphs.  I
  am grateful for long critical discussions with L.Biferale, M.Cencini,
  M.De Pietro, A.Giuliani and V.Lucarini: they provided hints and
  stimulated the ideas here.}

\*\*

  \halign{&\small#\hfill\cr &Giovanni Gallavotti\cr &Universit\`a di Roma
    1, INFN, \cr &e-mail: {\tt giovanni.gallavotti@roma1.infn.it}\cr}

\bibliographystyle{apsrmp} 

\small \bibliographystyle{unsrt}

\end{document}